\documentstyle[12pt,aaspp4]{article}
\begin{document}

\title{Keck Speckle Imaging of the White Dwarf G29-38: No Brown Dwarf Companion Detected}
\author{Marc J. Kuchner, Christopher D. Koresko, and Michael E. Brown\altaffilmark{1}}
\affil{Division of Geological and Planetary Sciences, California Institute of Technology, Pasadena, CA 91125}
\altaffiltext{1}{Alfred P. Sloan Research Fellow}
\authoremail{mjk@gps.caltech.edu}

\begin{abstract}
The white dwarf Giclas 29-38 has attracted much attention
due to its large infrared excess and the suggestion that excess
might be due to a companion brown dwarf.  We observed this
object using speckle interferometry at the Keck telescope, obtaining diffraction-limited resolution (55 milliarcseconds)
at K band, and found it unresolved.  Assuming the entire K band excess is
due to a single point-like companion, we place an upper limit on the binary separation of $30$ milliarcseconds, or $0.42$ AU at the star's distance of 14.1 pc.  This result, combined with astroseismological data and other images of G29-38, supports the hypothesis that the source of the near-infrared excess is not a cool companion but a dust cloud.

\end{abstract}


\section {Introduction}

\markcite{zuck87} Zuckerman and Becklin (1987) discovered that the
white dwarf Giclas 29-38 has
a large infrared excess and proposed that
the excess could be due to a brown dwarf companion.  This suggestion
inspired discussion of brown dwarfs as white dwarf companions
\markcite{stri90} (Stringfellow, Black \& Bodenheimer 1990),
oscillating brown dwarfs \markcite{marl90} (Marley, Lunine \&
Hubbard 1990), and other possible cool companions
that could explain the excess \markcite{gree88} (Greenstein 1988).
Later photometry by \markcite{toku90}Tokunaga
et al. (1990) and \markcite{tele90}Telesco, Joy \& Sisk (1990)
suggested that the 10 micron excess greatly exceeds that
expected from a brown dwarf companion, leading to the interpretation that the mid-infrared excess originates from a cloud of circumstellar dust.
However, new data from ISOCAM \markcite{char98}(Chary, Zuckerman \& 
Becklin 1998) show that the 7 and 15 micron excesses are in agreement 
with a 1000 K blackbody fit to the excess at other wavelengths. 
The source of the infrared excess of G29-38 remains uncertain.

Direct searches for a companion have produced mixed results.  Tokunaga et al. \markcite{toku88}(1988) imaged G29-38 at H and K bands and limited the
extent of the source to a diameter of 400 milliarcseconds (mas) or 5.64 AU.  Tokunaga et al. \markcite{toku88}(1988) and Tokunaga et al.
\markcite{toku90}(1990) took near-infrared spectra of the object and
found no evidence for
absorption features due to a brown dwarf.  Haas and Leinert
\markcite{haas90}(1990) took slit scans of G29-38 in 1988, and
found a North-South extension at K-band that was well fit by a binary
model with a flux ratio of 1:1 and a separation of $230 \pm 40$ mas ($3.24 \pm 0.56$ AU).  However, when Haas and Lienert repeated their
observations the following year under better
seeing conditions, the object appeared unextended.  Shelton, Becklin
and Zuckerman \markcite{shel98}(1998) took slit scans of G29-38 in
the J and K bands at the Lick 3-meter telescope in October of 1989 to look
for the centroid shift that would arise if, as the photometry suggests,
the hypothetical cool companion is brighter in K and the white dwarf
is brighter in J.  They did not see this effect.  They place an upper limit of
$40$ mas (0.56 AU) on the North-South binary separation, and
an upper limit of $120$ mas (1.69 AU) on the East-West separation.

Attempts to find the radial velocity signature of a companion to G29-38
have also proven frustrating.  \markcite{barn92}Barnbaum \& Zuckerman (1992)
combined their own spectroscopy with radial velocity data by \markcite{grah90}Graham et al. (1990),
Graham, Reid, \& Rich (1991, personal communication reported in Graham
et al. 1990), Liebert \& Saffer (1989,
personal communication reported in Graham et al. 1990) and \markcite{lieb89}
Liebert, Saffer, \& Pilachowski (1989), and
reported a probable radial velocity variation with a period of
11.2 months and an amplitude of $10 \ {\rm km} \ {\rm s}^{-1}$. Kleinman et al.
\markcite{klei94}(1994), however, argued
based on extensive astroseismological observations that the radial velocity variation due to a binary companion must be less than $\pm 0.65 \ {\rm km} \ {\rm s}^{-1}$ assuming a $\sim$1 year period.   

Hoping to find another clue to the mystery of the infrared excess,
we imaged G29-38 at K band on the 10-m W. M. Keck
telescope using speckle interferometry to search for a resolved companion
at the diffraction limit. 

\section {Observations}

We imaged G29-38 at K band with NIRC (the Near-Infrared Camera; \markcite{matt94}Matthews \& Soifer 1994) on the W. M. Keck telescope
on December 15, 1997.  The seeing was extraordinary; we used 0.5 second integrations and saw about 5 speckles and a diffraction-limited core. 
We took 12 sets of 100 frames of G29-38.  Among observations of
G29-38 we interspersed observations of two nearby, presumably unresolved
calibrator stars, S23291+0515 and S23292+0521, which we observed in the
same manner as G29-38, for a total of 6 sets of calibrator frames.  We
used a version of the speckle reduction software described in Koresko et al. \markcite{kore91}(1991) adapted for use with NIRC.  We chose
a $128 \times 128$ pixel subframe centered on the object, and constructed
$128 \times 128$ pixel sky frames from the corners of the
$256 \times 256$ pixel NIRC images.  From each set of object and sky frames
we computed a power spectrum, and a bi-spectrum, and
re-constructed Fourier phases and amplitudes.  We divided the Fourier components
from each target set by the Fourier components from a few different calibrator
sets to correct for the telescope-aperture transfer function, and in
this way assembled 18 calibrated images and 18 calibrated power spectra.

Figure 1 shows the mean of the images, compared to a simulated image of a
point source---the Fourier transform of the Gaussian$\times$Hanning apodizing function used to synthesize the speckle images.  The plate scale is 20.57 mas per pixel.  Figure 2 shows the azimuthal average of the arithmetic
mean of the calibrated power spectra, where we normalized each power
spectrum by dividing it by the geometric mean of the first 15 data points after
the zero-frequency component.
The error bars represent the 1-$\sigma$ variations among the 18 power spectra.  The $\lambda/D$ diffraction limit of Keck at K-band is 55 mas.  The noise increases at high frequencies because the power in the images decreases
near the diffraction limit.  The low frequency spike probably occurs because of
seeing noise, the change in seeing between observations of G29-38 and observations of the calibrators.  Because the final image closely resembles a point source and the power spectrum is consistent with a constant,
the power spectrum of a $\delta$-function, we conclude that we did not resolve G29-38.

\section {Discussion}

The K-band flux of G29-38 is $5.46 \pm 0.15$ mJy; 2.05 mJy of this is in excess of Greenstein's \markcite{gree88}(1988) white dwarf model
\markcite{toku90}(Tokunaga, Becklin \& Zuckerman 1990).
We computed the power spectrum of a binary system consisting of
a Greenstein white dwarf and a point-like companion which supplies
all the excess flux.  The only free parameter for this binary model is the angular separation of the components.  We fit the model to the observed power spectrum, and derive a best fit binary separation of 20 mas.  The maximum deviation of the power spectrum from a straight line, however, is consistent with typical deviations due to time variations of the atmosphere-telescope point-spread function.  In figure 2, we compare the 20 mas model with the observed power spectrum and a model with the same flux ratio but a 30 mas separation.  The latter model is marginally inconsistent with our observations, so we report 30 mas as an upper limit to the binary separation.

At G29-38's distance of 14.1 pc \markcite{toku90}(Tokunaga et al. 1990), 30 mas corresponds to a transverse separation
of 0.42 AU.  Assuming that G29-38 is 0.61 $M_{\odot}$
\markcite{berg95}(Bergeron et al. 1995), an 0.06 $M_{\odot}$
brown dwarf orbiting the star at 0.42 AU would have a period of about
0.33 years and would create a reflex motion in G29-38 that
would have been detectable to Kleinman et al. \markcite{klei94}(1994)
if the orbit were inclined more than 10 degrees from face-on.  The statistical likelyhood of an inclination $\le 10$ degrees is 1.5\%.  Closer orbits
would be easier to detect from reflex motion.

Perhaps a brown dwarf orbits G29-38 with a long period that would be hard to identify in reflex motion and the brown dwarf happened to pass in front of the star or behind it when we observed it on December 15, 1997.  For instance, Kleinman et al. \markcite{klei94}(1994) saw a long-term trend in their radial velocity data which could be interpreted as a companion with an $\sim 8$ year period causing radial velocity variations on the order of 0.8 km/s.  Such a companion would have a semimajor axis of $\sim 3.4$ AU.   If the orbit had a semi-major axis $a$, and were edge-on, the fraction of the time the brown dwarf would spend in the region where we couldn't resolve it is $\sim {2 \over \pi} \sin^{-1}{0.42 AU \over a}$; for $a=3.4$ AU, there is a $<8$\% chance that the brown dwarf would have been hidden from us.  Since Shelton et al. \markcite{shel98}(1998) also missed the hypothetical edge-on brown dwarf in 1989 as it passed close to the star, we find this scenario unlikely.

A companion in an eccentric orbit is easier to detect from reflex motion than a companion in a circular orbit with the same semi-major axis.  Therefore such a companion would have to be farther away from the star on average for Kleinman et al. \markcite{klei94}(1994) to have missed it, making it even more unlikely that it would have been hidden from us, Haas \& Lienert \markcite{1990}(1990), and Shelton et al. \markcite{shel98}(1998).  A companion in an eccentric, face-on orbit would spend relatively little time close to the star, and probably would not have been missed by both us and Shelton et al. \markcite{shel98}(1998). 

The infrared excess may represent thermal radiation from a cloud of dust rather than a cool companion \markcite{zuck87}(Zuckerman
\& Becklin 1987).  We can place no constraints on the concentration or geometry of such a cloud.  Dust radiating thermally at 1--15 microns heated by radiation from the white dwarf alone would be far too close to the star ($ < 10^{-3}$ AU) for us to resolve. 

\section {Conclusions}

We conclude that the infrared excess of G29-38 is not due to a single
orbiting companion.  If there were a single companion producing the excess, it would have to orbit almost face-on and closer than 0.4 AU;  or it could orbit roughly edge on, with a period of several years, in such a way that it happened to appear at a minimum angular separation from the star in December, 1997 when we observed it and in the fall of 1989 when Haas \&
Lienert \markcite{haas90}(1990) and Shelton et al. \markcite{shel98}(1998) observed it.  Either case is highly improbable.  This result supports the hypothesis that source of the near-infrared excess is not
a cool companion but a dust cloud \markcite{zuck87}(Zuckerman
\& Becklin 1987; \markcite{wick87}Wickramasinghe et al. 1987; \markcite{grah90} Graham et al. 1990; \markcite{koes97} Koester et al. 1997).

\acknowledgments

We thank Eugene Chiang, Chris
Clemens, Peter Goldreich, and Ben Zuckerman for inspiration and
helpful discussions.  The observations reported here were obtained at
the W. M. Keck Observatory, which is operated by the California
Association for Research in Astronomy, a scientific partnership among
California Institute of Technology, the University of California, and
the National Aeronautics and Space Administration.  It was made possible
by the generous financial support of the W. M. Keck Foundation.

\newpage

\figcaption{Reconstructed K-band speckle image of G29-38 compared to a synthesized image of a point source.  Both are normalized so their intensities range from 0 to 1, and have contour levels of 0.1, 0.3, 0.5, 0.7, and 0.9.
G29-38 is unresolved. \label{fig1}}

\figcaption{Azimuthally averaged spatial power spectrum of G29-38 compared to simulated azimuthally-averaged power spectra of a binary with a flux ratio equal to the G29-38's K-band excess.  The error bars represent 1-$\sigma$ variations among the 18 object-calibrator pairs.  If G29-38 were a binary with this K-band flux ratio and the separation were larger than 30 mas, we would
have detected the companion.  \label{fig2}}

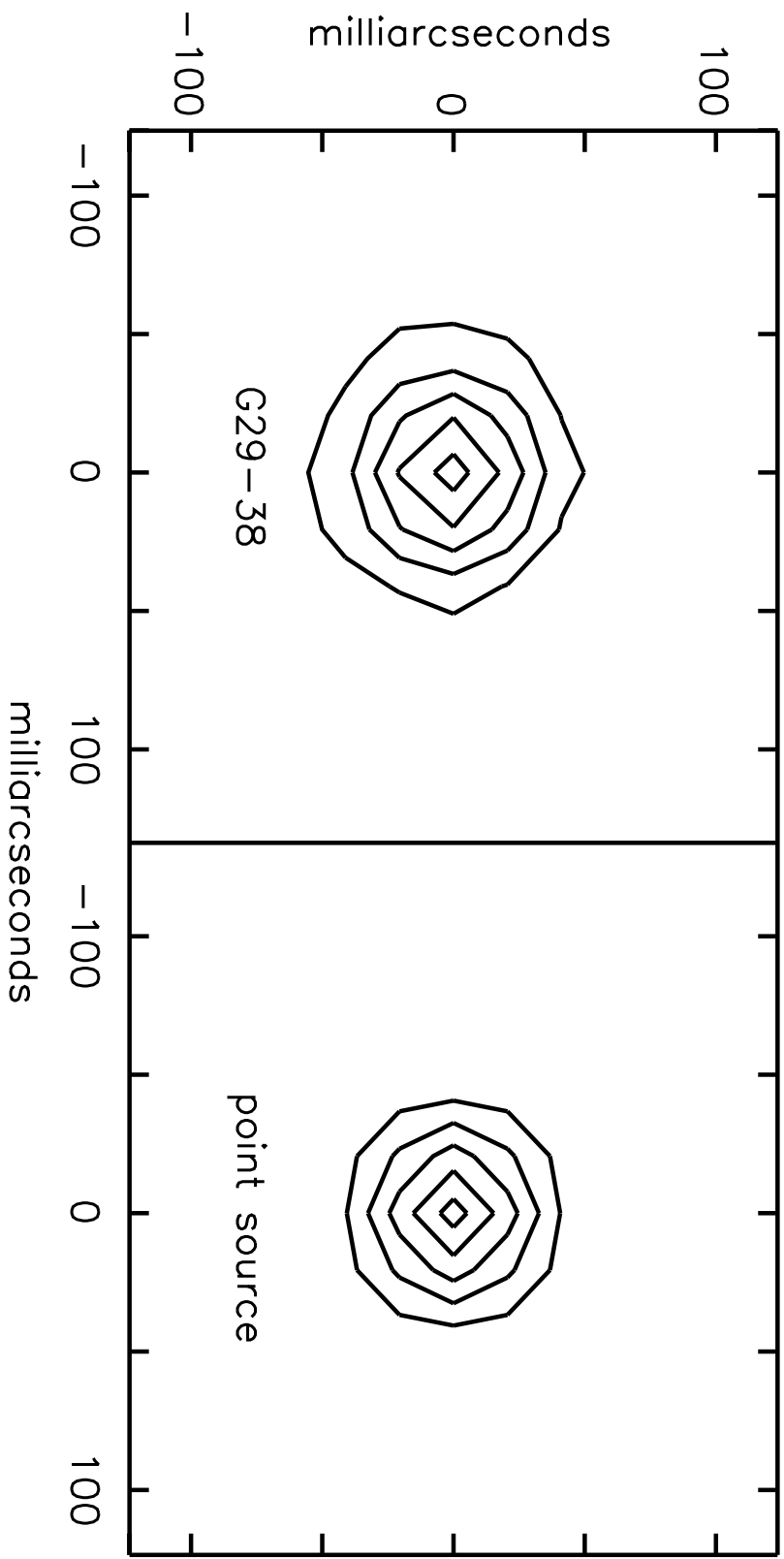
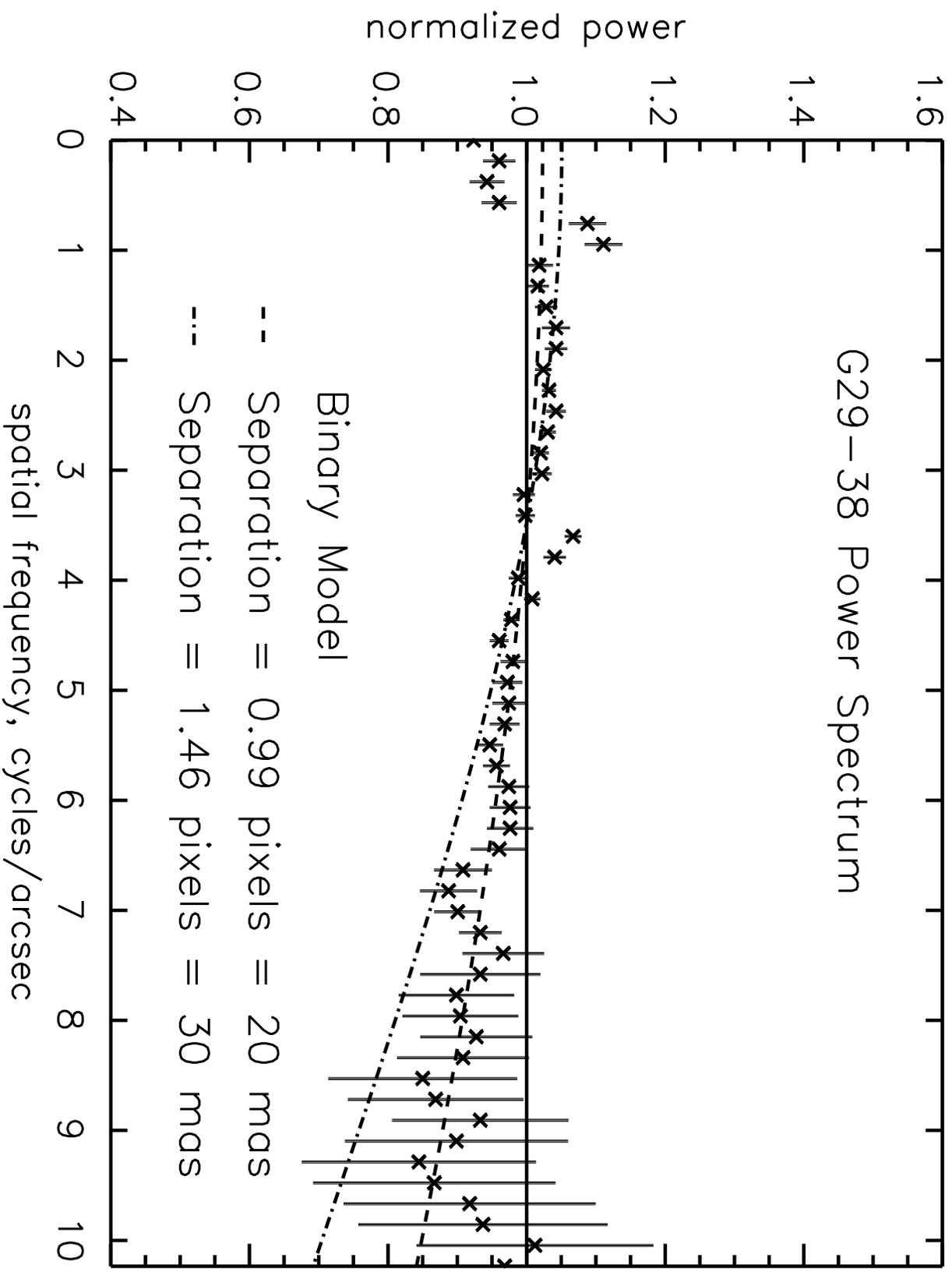
\end{document}